\documentclass[prl,floatfix,twocolumn,superscriptaddress]{revtex4-2}
\usepackage{amsmath}
\usepackage{amssymb}
\usepackage{amstext}
\usepackage{amsopn}
\usepackage{amsfonts}
\usepackage{amsxtra}
\usepackage[english]{babel}
\usepackage{graphicx}
\usepackage{float}
\usepackage{bm}
\usepackage{multirow}
\usepackage{dcolumn}
\usepackage{color}
\usepackage{hyperref}

\newcommand{\icm}{\ensuremath{~\textrm{cm}^{-1}}}

\begin{document}

\title{\textbf{Charge Dynamics of an Unconventional Three-Dimensional Charge Density Wave in Kagome FeGe}}

\author{Shaohui Yi}
\thanks{These authors contributed equally to this work.}
\author{Zhiyu Liao}
\thanks{These authors contributed equally to this work.}
\affiliation{Beijing National Laboratory for Condensed Matter Physics, Institute of Physics, Chinese Academy of Sciences, P.O. Box 603, Beijing 100190, China}
\affiliation{School of Physical Sciences, University of Chinese Academy of Sciences, Beijing 100049, China}

\author{Qi Wang}
\thanks{These authors contributed equally to this work.}
\affiliation{School of Physical Science and Technology, ShanghaiTech University, Shanghai 201210, China}
\affiliation{ShanghaiTech Laboratory for Topological Physics, ShanghaiTech University, Shanghai 201210, China}

\author{Haiyang Ma}
\author{Jianpeng Liu}
\affiliation{School of Physical Science and Technology, ShanghaiTech University, Shanghai 201210, China}

\author{Xiaokun Teng}
\author{Bin Gao}
\author{Pengcheng Dai}
\affiliation{Department of Physics and Astronomy, Rice University, Houston, Texas 77005, USA}

\author{Yaomin Dai}
\affiliation{National Laboratory of Solid State Microstructures and Department of Physics, Nanjing University, Nanjing 210093, China}

\author{Jianzhou Zhao}
\email[]{jzzhao@swust.edu.cn}
\affiliation{Co-Innovation Center for New Energetic Materials, Southwest University of
Science and Technology, Mianyang 621010 Sichuan, China}

\author{Yanpeng Qi}
\email[]{qiyp@shanghaitech.edu.cn}
\affiliation{School of Physical Science and Technology, ShanghaiTech University, Shanghai 201210, China}
\affiliation{ShanghaiTech Laboratory for Topological Physics, ShanghaiTech University, Shanghai 201210, China}
\affiliation{Shanghai Key Laboratory of High-resolution Electron Microscopy, ShanghaiTech University, Shanghai 201210, China}

\author{Bing Xu}
\email[]{bingxu@iphy.ac.cn}
\affiliation{Beijing National Laboratory for Condensed Matter Physics, Institute of Physics, Chinese Academy of Sciences, P.O. Box 603, Beijing 100190, China}
\affiliation{School of Physical Sciences, University of Chinese Academy of Sciences, Beijing 100049, China}

\author{Xianggang Qiu}
\email[]{xgqiu@iphy.ac.cn}
\affiliation{Beijing National Laboratory for Condensed Matter Physics, Institute of Physics, Chinese Academy of Sciences, P.O. Box 603, Beijing 100190, China}
\affiliation{School of Physical Sciences, University of Chinese Academy of Sciences, Beijing 100049, China}

\date{\today}
%
%

\begin{abstract}
We report on the charge dynamics of kagome FeGe, an antiferromagnet with a charge density wave (CDW) transition at $T_{\mathrm{CDW}} \simeq 105$ K, using polarized infrared spectroscopy and band structure calculations. We reveal pronounced optical anisotropy along the $a$- and $c$-axis, as well as an unusual response associated with three-dimensional CDW order. Above $T_{\mathrm{CDW}}$, there is a notable transfer of spectral weight (SW) from high to low energies, promoted by the magnetic splitting-induced shift in bands. Across the CDW transition, we observe a sudden SW transfer from low to high energies over a broad range, along with the emergence of new excitations around 1\,200\icm. These results contrast with observations from other kagome metals like CsV$_3$Sb$_5$, where the nesting of VHSs leads to a clear CDW gap feature. Instead, our findings can be accounted for by a $2\times2\times2$ CDW ground state driven by a first-order structural transition involving large partial Ge1-dimerization. Our study thus unveils a complex interplay among structure, magnetism, and charge order, offering valuable insights for a comprehensive understanding of CDW order in FeGe.
\end{abstract}


\maketitle

%
%

The kagome lattice, a hexagonal network of corner-sharing triangles, has been studied for over 70 years~\cite{Syozi1951}. Its unique band structure features the coexistence of flat bands (FBs), Dirac crossings, and van Hove singularities (VHSs), making it an excellent platform for studying the variety of emergent quantum phases resulting from the complex interplay between geometry, topology, and electronic correlations. In the early days, research mainly focused on the geometric spin frustration, showing its great potential to realize quantum spin liquid states~\cite{Balents2002PRB,Anderson1973MRB,Balents2010Nature,Yan2011SC}. Subsequently, a range of topological quantum states have been explored, such as Weyl fermions in Co$_3$Sn$_2$S$_2$~\cite{Liu2018NP,Liu2019Science,Morali2019SA}, Dirac fermions and flat bands in CoSn~\cite{Liu2020NC}, and Chern gapped Dirac fermions in TbMn$_6$Sn$_6$~\cite{Yin2020Nat}. More recently, charge density wave (CDW) and unconventional superconductivity~\cite{Ortiz2020PRL,Guguchia2023NC,Zhao2021Nat,Mielke2022Nat}, as well as other exotic quantum phenomena, including electronic nematicity~\cite{Nie2022Nat}, roton pair density wave~\cite{Chen2021Nat}, and giant anomalous Hall effect~\cite{Yang2020SA}, have been reported in the non-magnetic kagome metals AV$_3$Sb$_5$ ($A = $ Cs, K, Rb).

\begin{figure*}[tb]
\includegraphics[width=2\columnwidth]{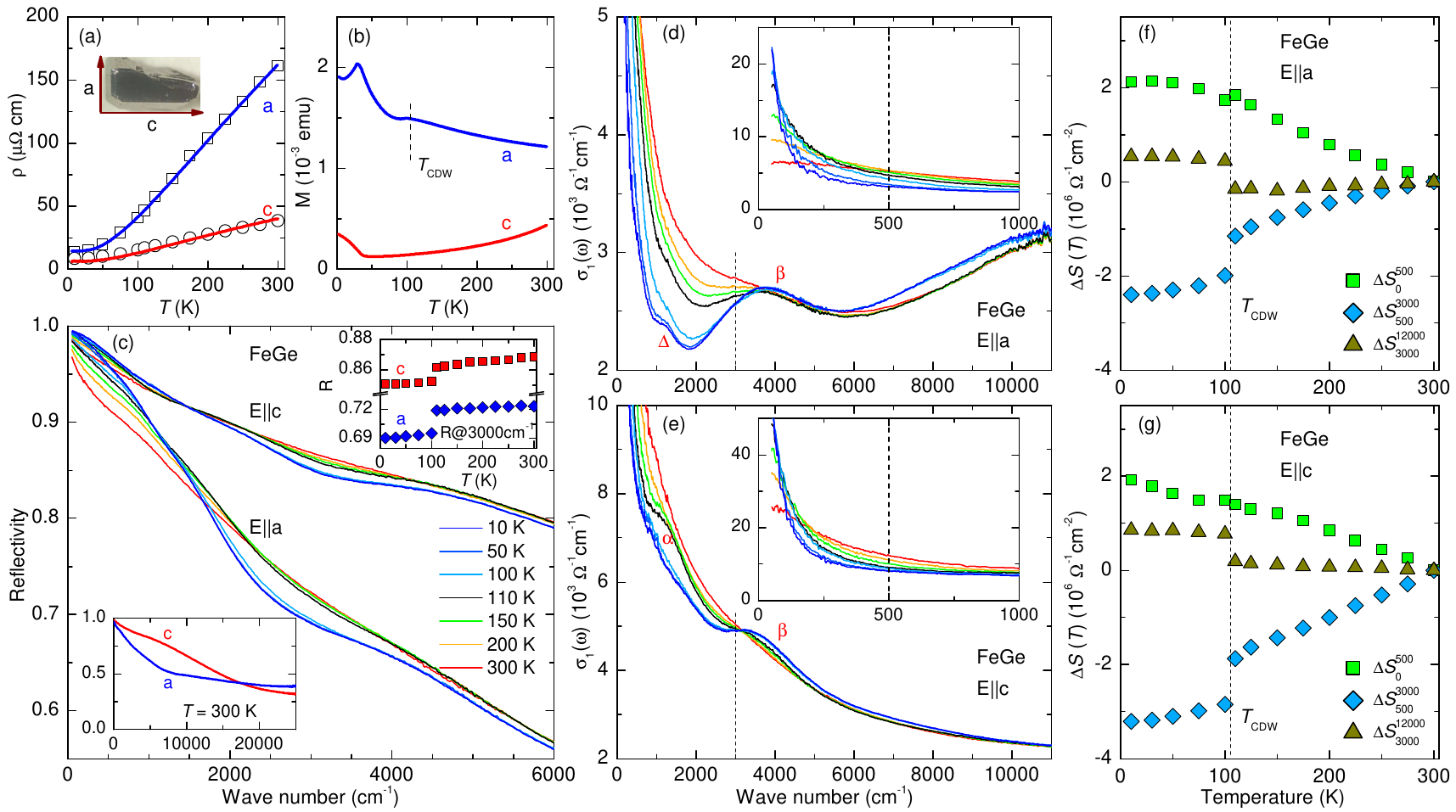}
\caption{(color online) (a) Temperature dependence of resistivity along the $a$-axis and $c$-axis for kagome FeGe. The open symbols represent $\rho \equiv 1/\sigma_1(\omega \rightarrow 0)$ obtained by the Drude fits to the optical data. (b) Temperature dependence of the magnetization curves $M_a(T)$ and $M_c(T)$ measured at a 1 T magnetic field. (c) Temperature-dependent spectra of reflectivity $R(\omega)$ for $\textbf{E} \parallel a$ and $\textbf{E} \parallel c$. Bottom inset: Spectra up to 25\,000\icm\ at 300~K. Top inset: Temperature dependence of $R(\omega = 3\,000\icm)$. (d) and (e) Temperature-dependent spectra of optical conductivity $\sigma_1(\omega)$ for $\textbf{E} \parallel a$ and $\textbf{E} \parallel c$, respectively. Insets provide the enlarged view of $\sigma_1(\omega)$ in the far-infrared region. (f) and (g) The corresponding changes of the spectral weight, $\Delta S_{\omega_a}^{\omega_b}(T) = S_{\omega_a}^{\omega_b}(T) - S_{\omega_a}^{\omega_b}(300 \mathrm{K})$, for $\textbf{E} \parallel a$ and $\textbf{E} \parallel c$, respectively.}
\label{Fig1}
\end{figure*}

Generally, the discovered kagome materials can only host either magnetism or charge orders, owing to the large energy separation between the FBs and the VHSs. However, a CDW order ($T_\mathrm{CDW} \simeq$ 100 K) has been found inside the antiferromagnetic (AFM) ordered phase of kagome FeGe ($T_N \simeq$ 410 K)~\cite{Teng2022Nat}. This CDW transition is associated with an increase of ordered magnetic moments~\cite{Teng2023NP}, which demonstrates an intertwined nature of magnetism and charge order (CO) in FeGe, thus offering a unique opportunity to explore a novel CDW with magnetism. Currently, the origin of CDW in FeGe is still full of controversy~\cite{Teng2022Nat,Teng2023NP,Shao2023ACSN,Wu2023CPL,Wang2023PRM,Zhao2023arXiv,WuSF2024PRX,Chen2024NC,
Shi2024,Wu2024PRL,Zhang2023arXiv,Yin2022PRL,Miao2023NC,Chen2023arXiv02,Ma2023arXiv,ZhouHJ2023PRB,Teng2024PRL,Klemm2024arXiv,Chen2024,Oh2024arXiv}. The nesting of VHSs at the M point and electron-phonon coupling were initially proposed to explain the formation of CDW~\cite{Teng2022Nat,Teng2023NP,Shao2023ACSN}, similar to AV$_3$Sb$_5$~\cite{Cho2021PRL,Tan2021PRL,Hu2022NC,Liu2021PRX,Kang2022NP,Zhou2021PRB,Luo2022NC,Xie2022PRB,Liu2022NC}. However, Wu \emph{et al.} found that the maximum nesting function of FeGe is at the K point instead of the M point, and suggested the key role of electronic correlations for CDW~\cite{Wu2023CPL}. Additionally, electronic correlations induces a softening effect along the L-H direction in the calculated phonon spectrum of FeGe~\cite{Miao2023NC,Chen2023arXiv02,Ma2023arXiv}. Furthermore, recent theoretical calculations and angle-resolved photoemission spectroscopy (ARPES) measurements in annealed samples support a new mechanism, in which the large dimerization partial Ge1-dimerization reduces the magnetic energy and leads to a stable $2 \times 2 \times 2$ CDW ground state~\cite{Wang2023PRM,Zhao2023arXiv,Oh2024arXiv}, in sharp contrast to AV$_3$Sb$_5$. Therefore, to clarify the origin of CDW in FeGe, a systematic study of charge dynamics across various electronic states or phases is essential.

In this Letter, we use polarized optical spectroscopy and density functional theory (DFT) calculations to systematically investigate the charge dynamics of the as-grown FeGe crystal across the CDW transition. Our study reveals a distinctive CDW response, characterized by the absence of a clear gap feature, a sudden SW transfer from low to high energies along both the $a$- and $c$-axis, and an emergent low-energy absorption. These observations suggest a first-order transition and three-dimensional (3D) nature for the CDW in FeGe, contrasting with the CDW behavior observed in AV$_3$Sb$_5$. Our experimental findings and the reproduced theoretical calculations support an unconventional CDW mechanism based on a first-order structural transition involving large partial Ge1-dimerization.

%

Sample synthesis, experimental methods, and details of Drude-Lorentz analysis and DFT calculations are provided in the Supplemental Material~\footnotemark[1].

%
%

Figure~\ref{Fig1}(a) displays the temperature ($T$) dependence of the resistivity for kagome FeGe along the $a$-axis (parallel to the kagome plane) and the $c$-axis (perpendicular to the kagome plane). Both directions exhibit typical metallic behavior, with strong electronic anisotropy characterized by lower resistivity along the $c$-axis. In Fig.~\ref{Fig1}(b), the $T$-dependent magnetizations, $M_a(T)$ and $M_c(T)$, show notable magnetic anisotropy. Upon cooling, $M_a(T)$ gradually increases and shows an anomaly at the CDW transition ($T_\mathrm{CDW} \simeq 105$ K). In contrast, $M_c(T)$ decreases and then turns upward below $T_\mathrm{Cant} \simeq 60$ K, attributed to spin canting~\cite{Bernhard1988JPF} or spin density wave order~\cite{Chen2024,Klemm2024arXiv}.

\begin{figure*}[tb]
\includegraphics[width=2\columnwidth]{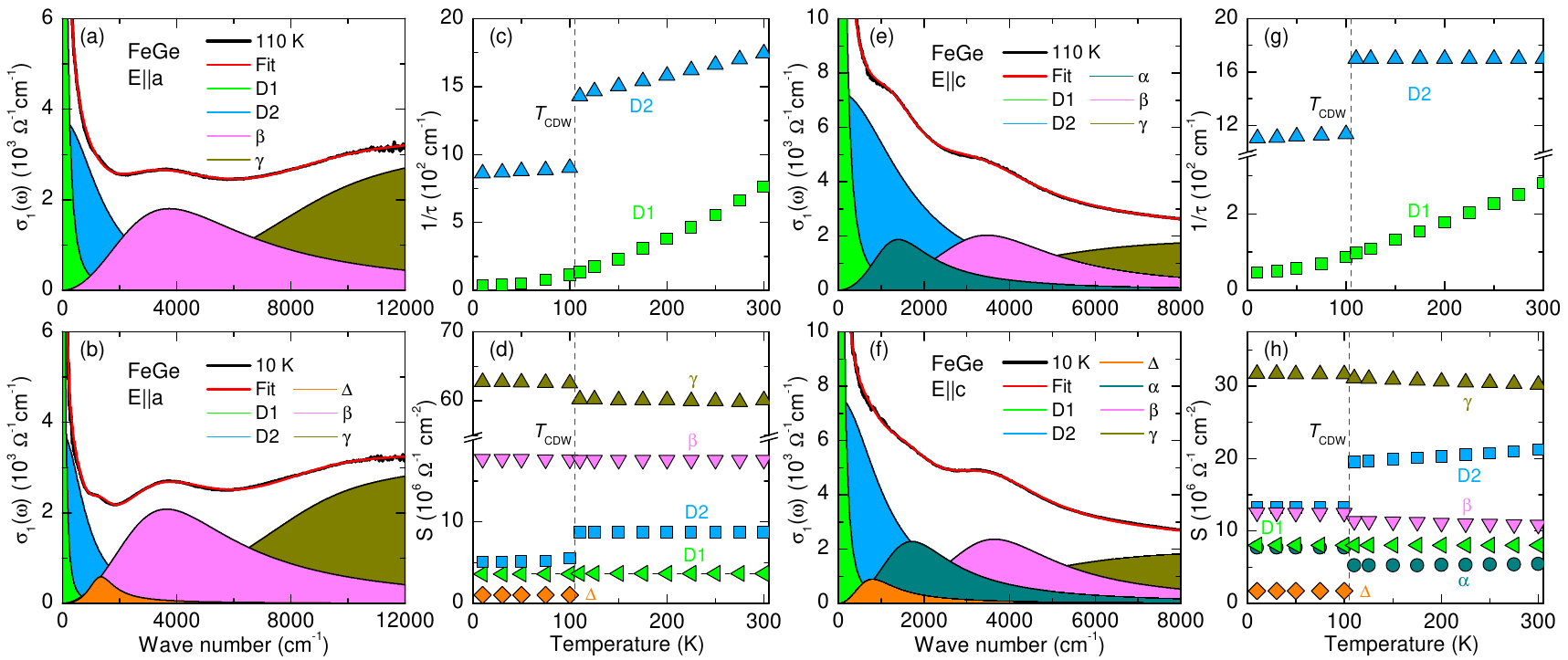}
\caption{(color online) (a) and (b) Decomposition of $\sigma_1(\omega)$ for $\textbf{E} \parallel a$ using a Drude-Lorentz model at $T = 110$ K and 10 K, respectively. (c) Temperature dependence of the scattering rate $1/\tau$ for the two Drude components. (d) Temperature-dependent spectral weight of various individual components. (e--h) Similar plots for $\textbf{E} \parallel c$.}
\label{Fig2}
\end{figure*}

Due to the electronic and magnetic anisotropies, we measured the polarized reflectivity $R(\omega)$ of FeGe. In Fig.~\ref{Fig1}(c), $R(\omega)$ is presented up to 6\,000\icm\ from 300 to 10 K for both $\textbf{E} \parallel a$ and $\textbf{E} \parallel c$. Additional data, including measurements with and without polarization, as well as high-temperature data across the AFM transition, are provided in the Supplemental Materials~\footnotemark[1]. $R_a(\omega)$ and $R_c(\omega)$ at $T = 300$ K are shown up to 25\,000\icm\ in the bottom inset of Fig.~\ref{Fig1}(c). In the infrared region, $R_c(\omega)$ is much higher than $R_a(\omega)$, indicating an optical anisotropy consistent with the lower resistivity along the $c$-axis. In the low-frequency limit, both $R_a(\omega)$ and $R_c(\omega)$ approach unity and increase with decreasing $T$, reflecting the metallic nature of FeGe. Moreover, below $T_\mathrm{CDW}$, $R(\omega)$ shows a sudden drop in the range of 2\,000 -- 5\,000\icm. Such a sudden change, as highlighted by the $T$-evolution of $R(\omega = 3000\icm)$ in the top inset, provides an initial spectroscopic indication of a first-order CDW transition in FeGe, which is consistent with the neutron~\cite{Teng2022Nat}, x-ray scattering~\cite{Miao2023NC}, and Raman experiments~\cite{WuSF2024PRX}.

The optical conductivity $\sigma_1(\omega)$ provides direct information about the charge dynamics. Fig.~\ref{Fig1}(d) and Fig.~\ref{Fig1}(e) display the $T$-dependent $\sigma^\mathrm{a}_{1}(\omega)$ and $\sigma^\mathrm{c}_{1}(\omega)$ for $\textbf{E} \parallel a$ and $\textbf{E} \parallel c$, respectively. In the far-infrared region, as highlighted in the insets, $\sigma_1(\omega)$ is characterized by a Drude-like peak at the origin. At high temperature, the Drude peak is quite broad, with a prominent tail extending toward higher frequencies, due to intraband excitations of carriers that are incoherent with a large scattering rate. As the temperature decreases, the Drude peak narrows significantly, with its low-frequency region becoming strongly enhanced while the high-frequency tail decreases accordingly, signaling a strong reduction in the scattering rate. These spectral changes thus reveals an evolution of incoherent excitations at high temperatures to coherent excitations at low temperatures. Such an incoherent-coherent behavior in optical conductivity has also been observed in iron-based superconductors~\cite{Wu2010,Dai2013PRL}. With the narrowing of the Drude response, $\sigma_1(\omega)$ in the mid-infrared region (500 -- 3\,000\icm) is suppressed and its spectral weight (SW) transfers to lower frequencies. Concurrently, an absorption peak (labeled as $\beta$) gradually emerges around 4\,000\icm. Across the CDW transition, $\sigma_{1}(\omega)$ below 3\,000\icm\ is further suppressed, while its associated SW is transferred to the $\beta$ peak and other high-energy interband transitions. The $T$-dependent spectral changes have been further analyzed using partial SW defined by $\Delta S_{\omega_a}^{\omega_b}(T) = S_{\omega_a}^{\omega_b}(T) - S_{\omega_a}^{\omega_b}(300~\mathrm{K})$, where $S_{\omega_a}^{\omega_b}(T) = \int_{\omega_a}^{\omega_b}\sigma_1(\omega, T)d\omega$. This approach allows us to identify the SW changes of different electronic excitations by selecting suitable cutoff frequencies $\omega_a$ and $\omega_b$. Fig.~\ref{Fig1}(f) and Fig.~\ref{Fig1}(g) detail the SW changes within different regions for $\textbf{E} \parallel a$ and $\textbf{E} \parallel c$, respectively. In the cases of $\Delta S_{0}^{500}$ and $\Delta S_{500}^{3000}$, the evolution from incoherent to coherent excitations results in an increase in $\Delta S_{0}^{500}$ and a corresponding decrease in $\Delta S_{500}^{3000}$. Meanwhile, $\Delta S_{3000}^{12000}$ remains almost $T$-independent at $T > T_{\mathrm{CDW}}$. For $T < T_{\mathrm{CDW}}$, both $\Delta S_{0}^{500}$ and $\Delta S_{500}^{3000}$ undergoes an abrupt suppression, coinciding with a significant increase in $\Delta S_{3000}^{12000}$, indicating a SW transfer over a broad energy scale due to the CDW transition.

\begin{figure*}[tb]
\includegraphics[width=2.02\columnwidth]{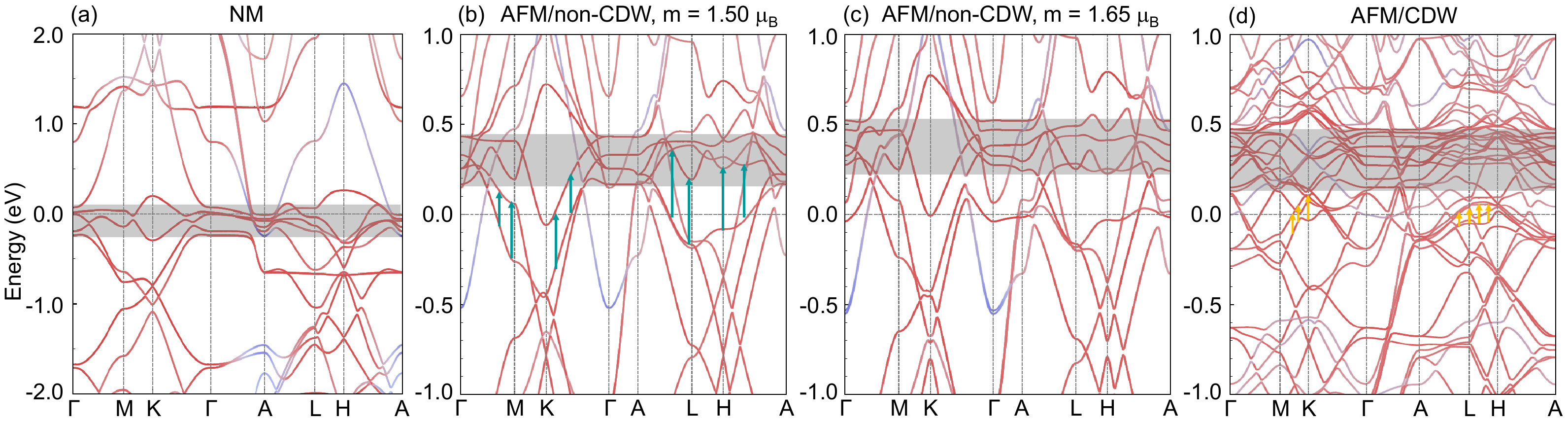}
\includegraphics[width=2\columnwidth]{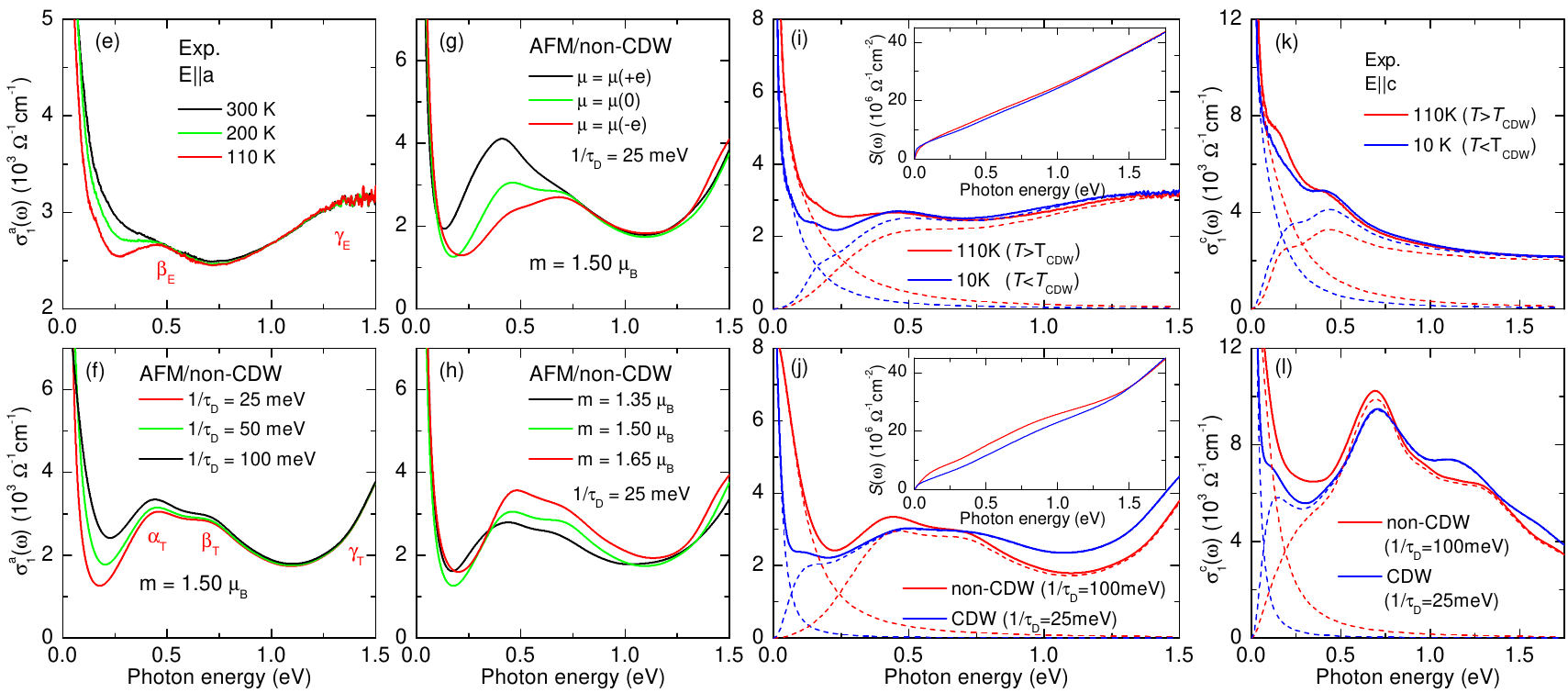}
\caption{(color online) Band structures of FeGe calculated in different phases: (a) non-magnetic (NM), (b--c) antiferromagnetic without CDW (AFM/non-CDW), and (d) antiferromagnetic with CDW (AFM/CDW). (e) Temperature-dependent $\sigma_1(\omega)$ measured with $\textbf{E} \parallel a$ above the CDW transition. The calculated $\sigma_1(\omega)$ in the AFM/non-CDW phase varies with (f) scattering rates, (g) electron doping levels, and (h) magnetic moments. A comparison of $\sigma_1(\omega)$ with and without the CDW obtained from (i) experimental measurements and (j) theoretical calculations. The dashed curves indicate the decomposition of intraband and interband excitations. Insets in (i) and (j) show the frequency-dependent spectral weight, $S(\omega)=\int^{\omega}_0\sigma(\omega^\prime)d\omega^\prime$, for both cases with and without the CDW. (k) and (l) show similar plots for $\sigma_1(\omega)$ with $\textbf{E} \parallel c$, for experimental and theoretical data, respectively.}
\label{Fig3}
\end{figure*}

Next, we employed the Drude-Lorentz model to fit the measured $\sigma_1(\omega)$ spectra. Fig.~\ref{Fig2}(a) and Fig.~\ref{Fig2}(b) show the decomposition of $\sigma^\mathrm{a}_{1}(\omega)$ at $T =$ 110 K and 10 K, respectively. It confirms that $\sigma^\mathrm{a}_{1}(\omega)$ above $T_\mathrm{CDW}$ can be described by two Drude components (a narrow one D1 in green and a broad one D2 in blue) and two lorentz components ($\beta$ in magenta and $\gamma$ in dark yellow). Below $T_\mathrm{CDW}$, an additional Lorentz band ($\Delta$ in orange) is required to account for the low-energy peak emerging around 1\,200\icm\ in the CDW state. Fits of $\sigma^\mathrm{a}_{1}(\omega)$ across all measured temperatures allow for the determination of the dc resistivity, $\rho \equiv 1/\sigma_1(\omega \rightarrow 0)$, shown by open symbols in Fig.~\ref{Fig1}(a), which aligns well with the dc transport measurement results. Fig.~\ref{Fig2}(c) and Fig.~\ref{Fig2}(d) display the temperature dependence of the fitting parameters, where the CDW transition strongly suppresses both the scattering rate and SW of D2 while enhancing the SW of $\gamma$ and $\Delta$. To minimize the number of the fitting parameters, the SW of each component was fixed at $T =$ 110 K and 10 K for $T > T_\mathrm{CDW}$ and $T < T_\mathrm{CDW}$, respectively. Notably, an equally good fit can be obtained by redistributing SW between D1 and D2, or between $\beta$ and $\gamma$.

A similar decomposition of $\sigma^\mathrm{c}_{1}(\omega)$ is shown in Figs.~\ref{Fig2}(e--h), where the single $\beta$ peak splits into two peaks ($\alpha$ and $\beta$). Below $T_{\mathrm{CDW}}$, the emergence of the $\Delta$ peak near the $\alpha$ peak causes the peak feature to blur in $\sigma^\mathrm{c}_{1}(\omega)$. Overall, both $\sigma^\mathrm{a}_{1}(\omega)$ and $\sigma^\mathrm{c}_{1}(\omega)$ exhibit similar responses to the CDW transition, providing strong evidence for the formation of a 3D CDW state in FeGe, consistent with the $2 \times 2 \times 2$ CDW order observed from other experiments~\cite{Chen2024NC,Shi2024,Miao2023NC}.

To gain deeper insights into the $T$-evolution of $\sigma_1(\omega)$, we conducted DFT calculations across various phases. In the non-magnetic (NM) phase, shown in Fig.~\ref{Fig3}(a), the band structure reveals typical kagome bands along the $\Gamma$-M-K-$\Gamma$ direction, with FBs (marked by a grey bar) near the Fermi level $E_{\mathrm{F}}$, VHSs below $E_{\mathrm{F}}$ at the M point and Dirac crossings at the K point. In the AFM phase, shown in Fig.~\ref{Fig3}(b), the FBs shift upward above $E_{\rm F}$ and the VHSs move closer to $E_{\rm F}$. Further comparison in Fig.~\ref{Fig3}(c) suggests that this upward shift of the FBs correlates with an increasing magnetic moment. In other words, the AFM order induces a transformation of the bands at $E_{\rm F}$ from flat to highly dispersive (e.g., the bands crossing $E_{\rm F}$ along the $\Gamma$-M direction), which accounts for the evolution of Drude response from incoherent to coherent intraband excitations. As compared in Fig.~\ref{Fig3}(e) and Fig.~\ref{Fig3}(f), the calculated $\sigma^\mathrm{a}_1(\omega)$ in the AFM/non-CDW phase captures all main experimental features. However, the calculated $\sigma^\mathrm{a}_1(\omega)$ reveals a nearby double peak ($\alpha_T$ and $\beta_T$) rather than the single peak ($\beta_E$) seen in experiments, with all peaks positioned at higher energies. This discrepancy arises from the absence of electronic correlations in the DFT calculations. In practice, electronic correlations would narrow the bands, leading the $\alpha_T$ and $\beta_T$ peaks to merge and shifting the absorption peaks to lower energies in actual measured data. Specifically, as detailed in the Supplemental Materials~\footnotemark[1], the shift in peak positions suggests a band renormalization factor of $1.6 \pm 0.2$, which closely aligns with the values observed in recent ARPES measurements~\cite{Teng2023NP} and dynamical mean-field theory (DMFT) calculations~\cite{Setty2022arXiv,Miao2023NC}, suggesting a moderate level of electronic correlations in FeGe.

Fig.\ref{Fig3}(g) and Fig.\ref{Fig3}(h) provide further calculations to demonstrate the impact of shifting the chemical potential ($\mu$) and varying the magnetic moment, respectively. Modifying $\mu$ by adding or removing one electron per unit cell influences only the SW of the $\alpha_T$ peak, as this peak is primarily governed by interband transitions close to $E_F$, indicated by the cyan arrows in Fig.\ref{Fig3}(b). An increase in the magnetic moment amplifies the SW of the $\alpha$, $\beta$, and $\gamma$ peaks. This contrasts with the reduction in the scattering rate of the Drude response, as simulated in Fig.\ref{Fig3}(f), where $\sigma_1(\omega)$ is suppressed in the region of high-energy interband transitions. Therefore, the spectral changes observed above $T_{\mathrm{CDW}}$, characterized by a narrowing Drude response and a nearly $T$-independent $\sigma_1(\omega)$ in the high-energy region, can be attributed to the combined effects of reduced scattering rates and band shifts due to gradually enhanced magnetic ordering with decreasing temperature.

The charge response to the CDW transition in FeGe is notably different from that of previous kagome metals, such as CsV$_3$Sb$_5$. As compared in the Supplemental Materials~\footnotemark[1], CsV$_3$Sb$_5$ exhibits a distinct CDW gap feature attributed to the nesting of VHSs around the M-point~\cite{Zhou2021PRB}, whereas FeGe shows no such gap feature. These experimental observations point to a different origin of the CDW in FeGe. Recent theoretical calculations suggest a novel CDW mechanism in which a $2 \times 2 \times 2$ CDW ground state is driven by a first-order structural transition involving large partial Ge1-dimerization~\cite{Wang2023PRM}. Following this mechanism, we constructed a $2\times2\times2$ superlattice to calculate the band structure and optical conductivity in the CDW state. As shown in Fig.~\ref{Fig3}(i) and Fig.~\ref{Fig3}(j), the calculated $\sigma_1(\omega)$ reproduces all experimental findings, including an additional low-energy peak around 0.15 eV and a SW transfer from low to high energies. The energy-dependent SW in the inset shows that this SW transfer occurs over a broad energy scale up to 1.5 eV. The band structure in the CDW phase, shown in Fig.~\ref{Fig3}(d), reveals two main changes. First, a strongly dispersive band near the $\Gamma$ point, shown in blue and primarily contributed by Ge-4p orbitals, narrows and shifts toward $E_F$ due to the in-plane Kekul\'{e}-type distortions and dimerization among the Ge atoms~\cite{Miao2023NC,Wang2023PRM,Zhang2024PRB}. Second, a denser set of bands appears near $E_F$, resulting from the band folding in the CDW state. The first change reduces the SW of the Drude response, while the second accounts for the additional low-energy excitations, as indicated by the orange arrows in Fig.~\ref{Fig3}(d). Furthermore, the ordered magnetic moment in the $2 \times 2 \times 2$ CDW ground state is enhanced via the partial Ge1-dimerization~\cite{Wang2023PRM,Teng2022Nat}, leading to an increased SW in the high-energy region, as also discussed in Fig.~\ref{Fig3}(h). These spectral changes occurring in a broad energy scale in FeGe are reminiscent of those observed in iron-based superconductors due to the effect of Hund's coupling~\cite{Wang2012,Schafgans2012}, as the enhanced spin polarization in the CDW state may lead to a stronger Hund's coupling. Additionally, a similar analysis of the results along the $c$-axis is plotted in Figs.~\ref{Fig3}(k--l). Overall, our experimental findings on the charge dynamics of the CDW transition are fully consistent with theoretical calculations, thereby supporting the novel CDW mechanism.

%
%
In summary, our optical conductivity measurements revealed rich information about the charge dynamics in FeGe, including a remarkable optical anisotropy, moderate electronic correlations, unconventional SW redistributions associated with magnetization-induced band shift and CDW transition, as well as a first-order transition and 3D character of CDW. These findings contrast with the conventional CDW mechanism observed in other kagome metals, and instead highlight a novel mechanism involving the intricate interplay among structure, magnetism, and charge order in FeGe.

%
%
\begin{acknowledgments}
We acknowledge discussions with Yilin Wang and Kai Wang. This work was supported by the National Key Research and Development Program of China (Grants Nos. 2022YFA1403900, 2024YFA1408301, 2023YFA1406002, and 2023YFA1607400), the National Natural Science Foundation of China (Grants Nos. 12274442, 12374155, and 52272265) and the Chinese Academy of Sciences (Grant No. XDB33010100). The single crystal synthesis and characterization work at Rice are supported by US NSF DMR-2401084 and the Robert A. Welch Foundation under Grant No. C-1839, respectively.
\end{acknowledgments}

\emph{Note added.--} During the revision of our manuscript, we noticed work that overlaps with some of our results~\cite{Wenzel2024PRL}.
%
%

\end{document}